\documentclass[twocolumn,trackchanges]{aastex63}

\shorttitle{Pickup ion acceleration at an oblique termination shock}
\shortauthors{Matsukiyo \& Matsumoto}

\graphicspath{{./}{figures/}}

\begin{document}


\title{Injection process of pickup ion acceleration at an oblique heliospheric termination shock}

\correspondingauthor{Shuichi Matsukiyo}
\email{matsukiy@esst.kyushu-u.ac.jp}

\author[0000-0002-4784-0301]{Shuichi Matsukiyo}
\affiliation{Faculty of Engineering Sciences, Kyushu University \\
6-1 Kasuga-Koen, Kasuga, Fukuoka 816-8580, Japan}
\affiliation{International Research Center for Space and Planetary Environmental Science (i-SPES), Kyushu University\\
Motooka, Nishi-Ku, Fukuoka 819-0395, Japan}
\affiliation{Institute of Laser Engineering, 
Osaka University \\
2-6, Yamadaoka, Suita, Osaka 565-0871, Japan}

\author[0000-0002-1484-7056]{Yosuke Matsumoto}
\affiliation{Institute for Advanced Academic Research, Chiba University, 1-33 Yayoi-cho, Inage-ku, Chiba 263-8522, Japan}


\begin{abstract}
Injection process of pickup ion acceleration at 
a heliospheric termination shock is investigated. 
Using two-dimensional fully kinetic particle-in-cell simulation, 
accelerated pickup ions are self-consistently reproduced 
by tracking long time evolution of shock with unprecedentedly 
large system size in the shock normal direction. 
Reflected pickup ions drive upstream large amplitude waves 
through resonant instabilities. Convection of the 
large amplitude waves causes shock surface reformation and 
alters the downstream electromagnetic structure. 
A part of pickup ions are accelerated to tens of upstream flow energy 
in the time scale of $\sim 100$ times inverse ion gyro frequency. 
The initial acceleration occurs through shock surfing 
acceleration mechanism followed by shock drift acceleration 
mechanism. Large electrostatic potential 
accompanied by the upstream waves 
enables the shock surfing acceleration to occur.

\end{abstract}



\section{\label{sec:intro}Introduction}

A collisionless shock is ubiquitous in space. It is an energy 
converter formed in a supersonic plasma flow. 
A variety of explosive phenomena in 
space, stellar wind, astrophysical jet accompany collisionless shocks. 
One of the outstanding issues of collisionless shock physics is the 
mechanism of particle acceleration occurring around it. It is believed 
that cosmic rays are produced 
in a collisionless shock through the diffusive shock acceleration (DSA) 
mechanism (e.g., \cite{blandford87}). In order for the DSA mechanism to work, 
preaccelerated nonthermal particles have to exist near a shock front. 
However, the mechanism producing such nonthermal particles, which should 
be dominated by complex microstructures of local electromagnetic fields, 
has been an open question for a long time. This is called the injection 
problem \citep{balogh13,burgess15,amano22}.

The heliospheric termination shock (HTS) is thought to be an ideal 
laboratory to study the injection process, because the anomalous 
cosmic rays (ACRs) having typically several tens of mega electron 
volts (MeV) are believed 
to be accelerated there \citep{chalov06}. However, the Voyager spacecraft 
observed only little amount of ACRs in the heliosheath, downstream 
of the HTS (e.g., \cite{stone05}). The reason for this was inferred by 
\cite{mccomas06} as that efficient acceleration of ACRs 
occurs in the flank regions of the HTS, where geometrical condition 
of the local HTS is more suitable for particle acceleration than 
that of the HTS where the two Voyager spacecraft crossed.

On the other hand, \cite{giacalone21} recently performed 
hybrid (kinetic ions with charge neutralizing electron 
fluid) simulations and showed that the acceleration rate of 
low energy ($\lesssim 50$ keV) pickup ions (PUIs) is more or 
less independent of the position in the HTS. The nonthermal particles in this energy range are expected to have 
their global map obtained in the near future through the IMAP 
(Interstellar Mapping and Acceleration Probe) mission. 
In their simulation \citep{giacalone21}, turbulence in the solar wind is taken into 
account. Particles are expected to undergo scattering by the 
turbulent field, leading to some 
degree of acceleration, and there is a potential for these 
particles to be injected into the DSA 
process 
\citep{giacalone21, trotta21, pitna21, zank21, trotta22, nakanotani22, wang23}. 
Another expected effect is the tilting of local magnetic 
field line. 
In oblique shocks particles reflected by the shock travel upstream, exciting 
waves themselves and generating mildly accelerated particles injected 
into the DSA process. Such a situation may occur in some specific 
regions of termination shock and perhaps locally everywhere 
when solar wind is turbulent. In this scenario, the generation of 
mildly accelerated particles needs to occur near the shock, 
but the detailed mechanism is not well understood. This study 
aims to validate the latter scenario.

It is thought that initial acceleration sets in when a particle 
is reflected at the shock. In general shock potential strongly 
affects the reflection of PUIs. 
In a hybrid simulation, the electrostatic potential is typically 
derived from a generalized Ohm’s law, which includes a term 
proportional to the electron pressure gradient. Calculating 
this electron pressure gradient term requires assuming a 
specific equation of state for electrons. Consequently, 
the electrostatic field in a hybrid simulation becomes model-dependent. 
\cite{swisdak23} pointed out that fully kinetic 
particle-in-cell (PIC) simulation leads to 
higher fluxes and maximal energies of PUIs than hybrid simulation 
likely due to differences in the shock potential. 
Before the Voyager spacecraft crossed the HTS, the shock 
surfing acceleration (SSA) was widely accepted as a plausible 
mechanism of injection \citep{lee96,zank96}.  
However, the Voyager 
did not observe the expected amount of high energy particles 
when they crossed the HTS \citep{decker08}. 
This implies that the shock surfing acceleration mechanism did not work effectively either. 
The reason for that SSA does not work in the termination 
shock was explained by \cite{matsukiyo14} 
by using one-dimensional fully kinetic particle-in-cell (PIC) 
simulation of (quasi-)perpendicular shock that 
most of potential jump in a PUI mediated shock occurs in 
an extended foot produced by reflected PUIs so that the 
potential jump at a shock ramp is insufficiently small for 
the SSA mechanism to work. 
The shock potential in an oblique PUI 
mediated shock has not been extensively studied so far.

Although the ab-initio PIC simulation is 
useful to reproduce complex multiscale electromagnetic structures 
responsible for injection processes, it requires large numerical 
cost. That is why PIC simulation of a collisionless shock in 
a plasma containing PUIs has been limitted mainly to the cases with 
one spatial dimension (e.g., \cite{lee05,matsukiyo07,matsukiyo11,matsukiyo14,oka11,lembege16,lembege18,lembege20}). 
Two-dimensional PIC simulations including PUIs 
are first conducted by \cite{yang15}. They focused on the 
impact of PUIs on shock front nonstationarity of a perpendicular 
shock, $\Theta_{Bn} = 90^{\circ}$, and energy dissipation up to 
$t=8 \Omega^{-1}_i$, where $\Theta_{Bn}$ denotes the shock 
angle, the angle between shock normal and upstream magnetic field, 
and $\Omega_i$ is upstream ion 
cyclotron frequency. \cite{kumar18} performed longer 
simulation up to $t \sim 30 \Omega^{-1}_i$ for quasi-perpendicular 
shocks, $\Theta_{Bn} = 80^{\circ}$. They paid more attention to 
energy distribution of PUIs and solar wind ions (SWIs) for different 
upstream velocity distribution functions of PUIs. 
A little longer calculation up to $t \sim 40 \Omega^{-1}_i$ was done 
with $\Theta_{Bn}=70^{\circ}$ by \cite{swisdak23}. 
However, since 
the shock angle is close to or equal to perpendicular in the 
above previous studies, significant acceleration of PUIs 
are not reproduced. Indeed, when considering 
the average values, the shock angle of the termination shock 
is nearly perpendicular. However, as mentioned in the previous 
paragraph, it is easy to predict that the local shock angle 
fluctuates significantly due to the effects of solar wind 
turbulence and unsteady solar activity. This fluctuation 
could have important implications for the ion injection process.

In this study we focus the initial acceleration, which is often 
called injection process, of PUIs at an oblique HTS. The 
microstructures of a local oblique HTS with $\Theta_{Bn}=50^{\circ}$, and with $60^{\circ}$ and $70^{\circ}$ for comparison,
mediated by the presence of PUIs and their impact on PUI 
acceleration are discussed by performing two-dimensional full 
PIC simulation with unprecedentedly long simulation time 
($t_{max} = 125 \Omega^{-1}_i$).


\section{\label{sec:setting}PIC simulation}

\subsection{Settings}
The system size in $x-y$ simulation domain is  
$L_x \times L_y = 2,000 v_A/\Omega_{i} \times 40.96 v_A/\Omega_{i}$ 
which is divided by 
$200,000 \times 4,096$ grid points. Here, $v_A$ denotes 
upstream Alfv\'{e}n velocity. 
The so-called injection method is utilized to form a shock. An upstream 
homogeneous plasma consisting of solar wind electrons and ions as well 
as PUIs is continuously injected from the moving boundary at finite $x$, 
while the plasma and electromagnetic waves are reflected at the fixed 
rigid wall boundary at $x=0$. The simulation frame is a downstream 
plasma frame so that a shock propagates toward positive $x-$direction. 
Initially 20 particles 
per cell are distributed for each species. The distribution functions 
of the injected solar wind electrons and ions are a shifted-Maxwellian 
and that of the injected PUIs is a shifted-shell distribution with 
zero width of the shell velocity. The 
injection (or drift) 
speed is $-3.75 v_A$ so that the Alfv\'{e}n Mach number of 
the shock is $M_A \approx 5.5$ for $\Theta_{Bn}=50^{\circ}$, 
$M_A \approx 5.8$ for $\Theta_{Bn}=60^{\circ}$, and $M_A \approx 5.9$ for 
$\Theta_{Bn}=70^{\circ}$, respectively. Here, the upstream magnetic field is 
in the $x-y$ simulation plane. These values of the Mach number are close to 
the one estimated by \cite{li08} using Voyager 2 data, 
while is a little smaller than the value assumed in \cite{giacalone21}. 
The upstream electron beta is 
$\beta_e=0.25$ and the solar wind ion temperature is the same as 
that of electrons, $T_i=T_e$. The beta value is slightly 
larger than the estimates by \cite{li08} and the values assumed by \cite{giacalone21}, but considering the significant variations in 
solar wind temperature observed by Voyager 2 \citep{richardson08}, 
it falls within a realistic range. The ion to electron mass ratio is 
$m_i/m_e=100$ for both the SWIs and PUIs, the ratio of 
upstream electron plasma frequency 
to cyclotron frequency $\omega_{pe}/\Omega_e=4$, and the relative 
PUI density is 25\%, respectively. In the following, 
the run with $\Theta_{Bn}=50^{\circ}$ is mainly discussed. 
Hereafter, time is normalized to $\Omega^{-1}_i$, velocity to 
$v_A$, and distance is to $v_A / \Omega_i$, respectively.

\subsection{Overview}
%
\begin{figure}[ht]
\includegraphics[clip, width=1.0\columnwidth]{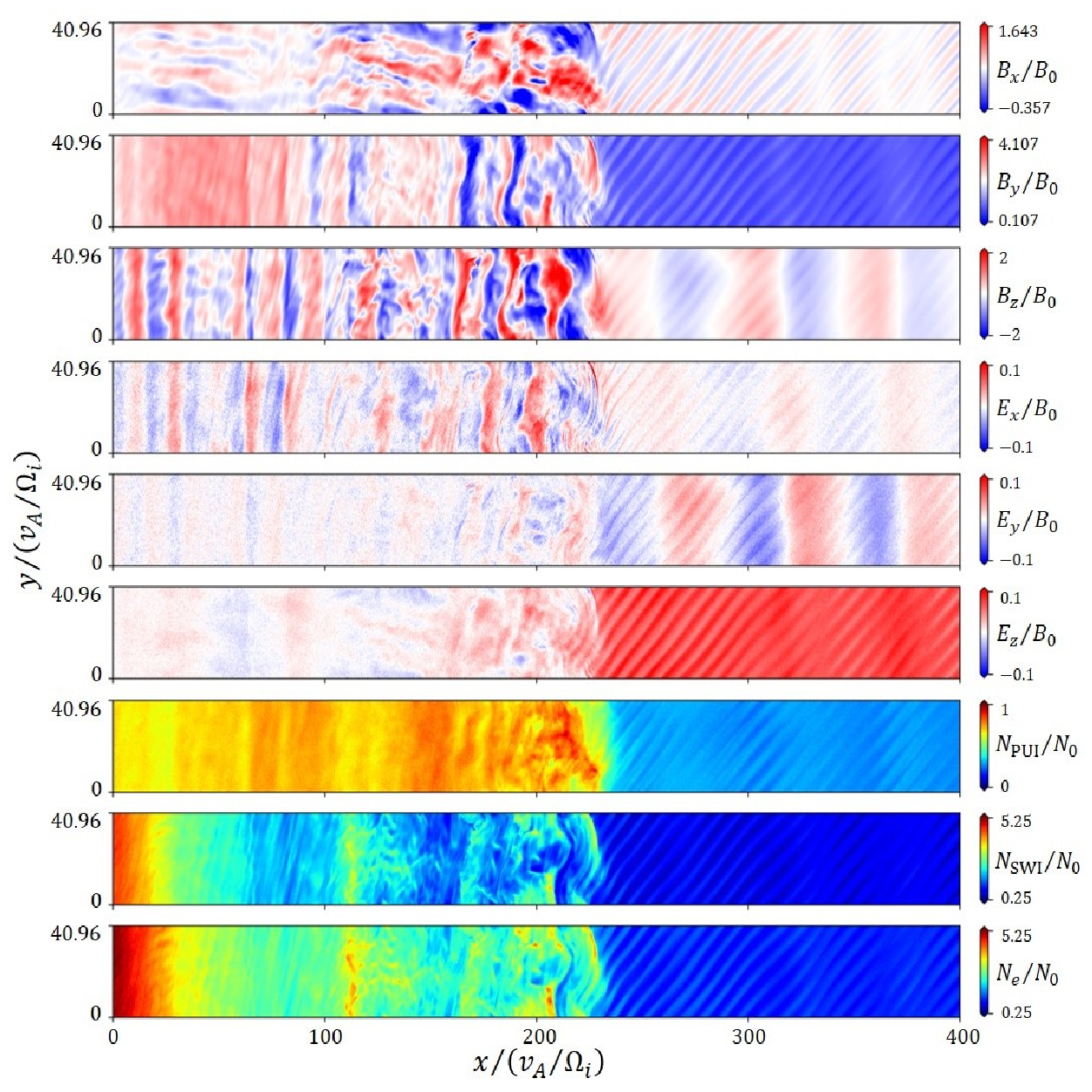}
\caption{\label{fig:fld_500000}Structure of a PUI mediated oblique 
shock at $t=125$ ($\Theta_{Bn}=50^{\circ}$). 
From the top, the first three panels denote 
magnetic field $B_x, B_y, B_z$, the next three panels electric field 
$E_x, E_y, E_z$, and the last three indicate density of PUIs 
($N_{\rm PUI}$), SWIs ($N_{\rm SWI}$), and electrons ($N_{e}$), 
respectively.}
\end{figure}
%
Fig.\ref{fig:fld_500000} shows the structure of a shock at 
$t = 125$, where a shock ramp is at $x \sim 210$. 
The first three panels represent magnetic field $B_x, B_y, B_z$, 
the next three electric field $E_x, E_y, E_z$, and the 
last three indicate density of PUIs ($N_{\rm PUI}$), SWIs 
($N_{\rm SWI}$), and electrons ($N_{e}$), respectively. Both 
upstream and downstream, large amplitude waves are generated. 

\cite{liu10} showed by performing a 2D hybrid simulation 
of a perpendicular shock including 20\% PUIs that Alfv\'{e}n 
ion cyclotron (AIC) instability and mirror instability lead 
to the downstream turbulent field. The AIC waves 
are visible in $B_x$ and $B_z$ and mainly propagate along 
the downstream magnetic field, while the mirror waves have 
wavenumbers 
oblique to it and are the most visible in $B_y$. In 
Fig.\ref{fig:fld_500000}, however, the AIC wave signature is 
confirmed only in $B_x$. The signature of mirror waves in 
$B_y$ and that of AIC waves in $B_z$ are overwhelmed by the 
vertical stripes which are also seen in $E_x$ and $E_y$. 

The upstream waves are dominated by two types, a relatively 
long wavelength waves propagating in negative $x-$direction 
(vertical stripes) and a relatively short wavelength waves 
propagating in almost perpendicular to the ambient magnetic 
field (oblique stripes). These waves are convected by the 
upstream flow. The former are not only the origin of the downstream 
vertical structures discussed above, but also the cause of 
shock surface reformation. We will discuss the 
generation mechanism of the upstream waves below.

\subsection{Upstream waves}

\begin{figure}[ht]
\includegraphics[clip, width=1.0\columnwidth]{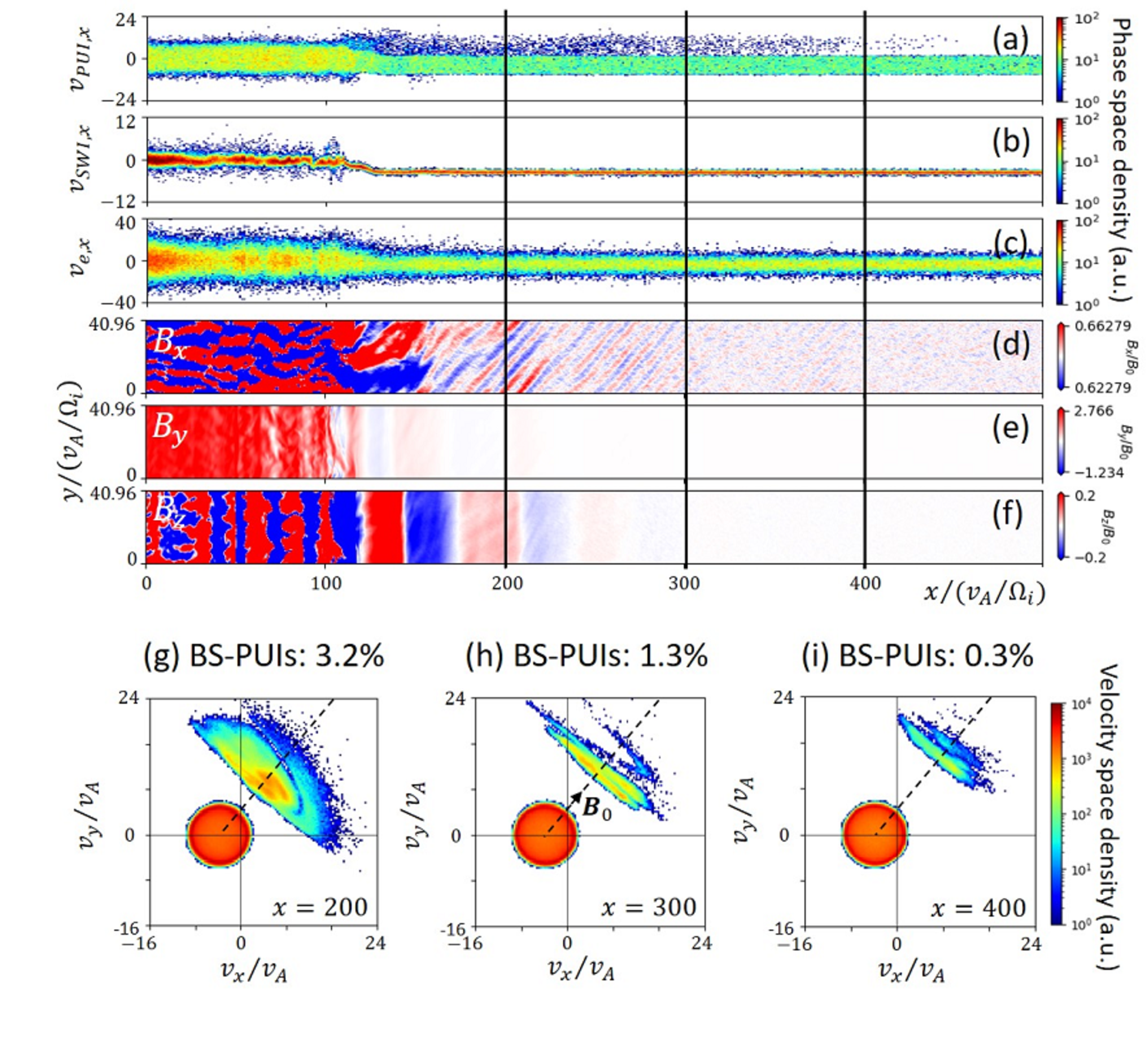}
\caption{\label{fig:pui_us_260000}Particle distribution and 
magnetic fields upstream of the shock at $t = 65$ 
($\Theta_{Bn}=50^{\circ}$). From 
the top, $V_x-x$ phase space distributions of (a) PUIs, (b) solar 
wind ions, (c) electrons, magnetic field components of (d) $B_x$, 
(e) $B_y$, (f) $B_z$, velocity distributions of PUIs at (g) 
$x=200$, (h) $x=300$, and (i) $x=400$, respectively. In (a-c) 
and (g-i), only the particles at $y \approx L_y/2$ are counted. 
The color bars in (a-c) denote phase space 
density (in an arbitrary 
unit). The color bar in (g-i) denote velocity space 
density (in an arbitrary 
unit). The dashed lines in (g-i) represent the direction 
of local (upstream) magnetic field, ${\bf B_0}$.
}

\end{figure}
%
In Fig.\ref{fig:pui_us_260000} the $V_x-x$ phase space distributions 
of (a) PUIs, (b) SWIs, and (c) electrons, whose positions 
in $y$ are $y \approx L_y/2$, are shown, the corresponding magnetic 
field structures are represented in (d-f), and velocity space 
distributions of the PUIs at different positions in $x$ are indicated 
in (g-i), respectively, at $t = 65$. 
The color scale in (a-c) and in 
(g-i) are in an arbitrary unit, and the color scale of 
the magnetic field is adjusted to appropriately see the upstream 
structures. It is clearly seen in (a) 
that some PUIs are reflected and backstreaming 
from the shock front at $x \approx 105$ toward upstream. 
On the other hand, as in the previous kinetic 
simulations of PUI mediated shocks \citep{matsukiyo14, wu09}, 
no SWIs are reflected (b). 
This implies that the upstream 
waves are generated by these backstreaming PUIs (BS-PUIs). 
In the local velocity space at (g) $x=200$, (h) $x=300$, and 
(i) $x=400$, the red circular populations denote incoming PUIs, 
while the BS-PUIs are indicated by the rest in each panel. 
The BS-PUIs are well separated in the velocity space from the 
incoming PUIs and their average velocities roughly align to 
the black dashed lines indicating the direction of upstream 
magnetic field. The relative density of 
the BS-PUIs is 0.3\% at $x=400$, 1.3\% at $x=300$, and 3.2\% at 
$x=200$, respectively.

Since the BS-PUIs are regarded as a field aligned beam, linear 
dispersion analysis of a beam-plasma system including background 
ions, electrons, and beam ions is performed. 
The details of the analysis are provided in the appendix.

In Fig.\ref{fig:disp} two unstable solutions obtained by 
solving the 
linearized Vlasov-Maxwell equation system are depicted. 
One is for $\theta_{Bk}=50^{\circ}$ and another is for 
$\theta_{Bk}=87^{\circ}$, where $\theta_{Bk}$ is 
the wave propagation angle with respect to the ambient magnetic 
field. For each solution, the solid line denotes the real part 
of frequency, while the dashed line shows the imaginary part. 
For $\theta_{Bk}=50^{\circ}$, the maximum growth rate occurs at 
$k \approx 0.096$. This corresponds to the relatively long 
wavelength waves having a wavenumber along the $x-$axis seen 
in Fig.\ref{fig:pui_us_260000}(e-f). The maximum growth rate 
occurs at $k \approx 1.2$ for $\theta_{Bk}=87^{\circ}$, 
which corresponds to the relatively short wavelength waves 
seen in Fig\ref{fig:pui_us_260000}(d). These instabilities are 
interpreted as a resonant instability with oblique propagation 
angle. 

%
\begin{figure}[ht]
\includegraphics[clip, width=1.0\columnwidth]{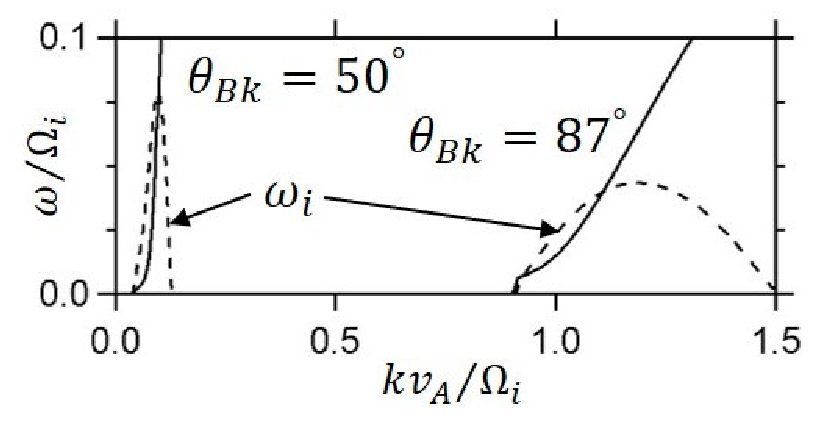}
\caption{\label{fig:disp}Linear dispersion relation of 
oblique resonant instability driven by BS-PUIs.}
\end{figure}
%

\subsection{Injection of PUIs}

\begin{figure}[ht]
\includegraphics[clip, width=1.0\columnwidth]{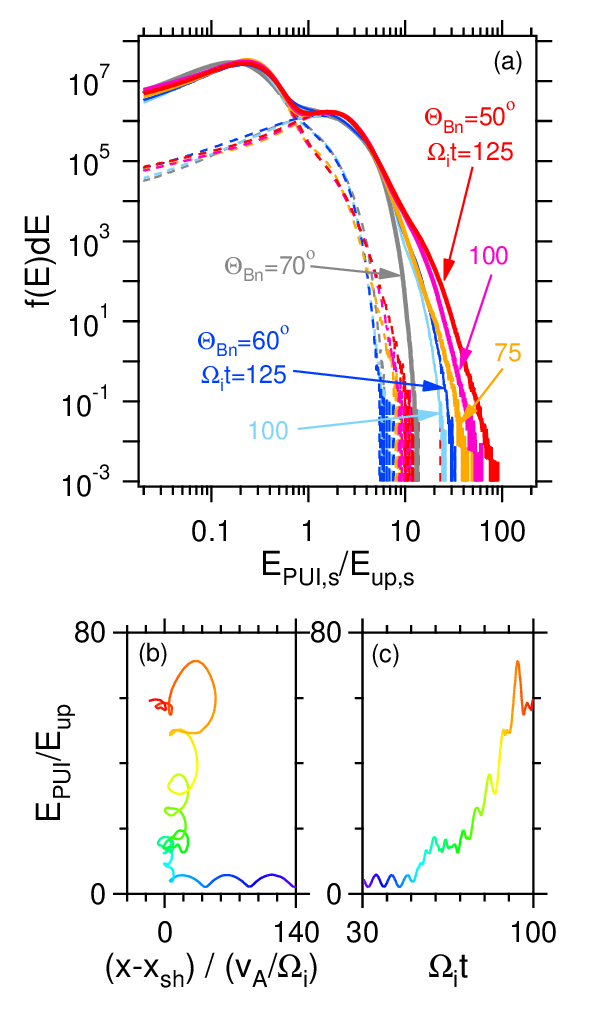}
\caption{\label{fig:PUIene} (a) 
Energy distributions of downstream PUIs at various shock 
angles and times. Each distribution (solid line) is divided 
into the two dashed lines indicating SWIs (lower energy) 
and PUIs (higher energy). 
Energy history of an accelerated PUI is shown 
by the rainbow 
colored lines as a function 
of (b) space and (c) time, respectively. The line color 
denotes time.}
\end{figure}
%
Fig.\ref{fig:PUIene}(a) shows downstream energy distribution 
functions of ions at different times, $t=75$ (orange), 100 (magenta), 
and 125 (red), where the energy is measured in the 
shock rest frame and normalized to the upstream 
bulk flow energy. 
The each solid line, indicating the distribution of whole ions, 
is divided into the two dashed lines with relatively low and 
high energy. The low and high energy ones denote solar 
wind ions and PUIs, respectively. 
It is 
evident that high energy tail of the PUIs evolves in time, 
indicating the system has not reached steady evolution state 
with a constant power-law index.

For comparison, simulations for $\Theta_{Bn}=70^{\circ}$ 
and $\Theta_{Bn}=60^{\circ}$ were additionally performed. 
At $\Theta_{Bn}=70^{\circ}$, no backstreaming ions were 
observed, and the downstream energy distribution remained 
steady during the same time interval, with no discernible 
accelerated component of PUIs. The energy distribution at 
this time ($t=125$) is shown by the gray line.  
At $\Theta_{Bn}=60^{\circ}$, a very small amount of backstreaming 
ions was observed. As indicated by blue and cyan lines in 
Fig.\ref{fig:PUIene}(a), the generation of accelerated PUIs can be 
observed over time, but the generation efficiency is evidently 
lower than that at $\Theta_{Bn}=50^{\circ}$. The energy 
densities of accelerated particles at $t=125$, relative to 
the steady downstream plasma energy density at 
$\Theta_{Bn}=70^{\circ}$, are 0.066 when $\Theta_{Bn}=60^{\circ}$ 
and 0.13 when $\Theta_{Bn}=50^{\circ}$, respectively. We consider 
the shock angle of $60^{\circ}$ to be close to the critical 
angle where particle backstreaming begins. At $\Theta_{Bn}=50^{\circ}$, 
the excitation of large-amplitude upstream waves by backstreaming 
PUIs is believed to increase the efficiency of accelerated PUI generation 
as discussed below.

The flux at the shoulder of the accelerated PUIs is 
roughly two orders of magnitude lower than the peak flux 
of PUIs. This relative flux of the shoulder looks comparable 
to or even higher than that seen in the past hybrid 
simulation with higher $M_A$ \citep{giacalone21}. 
To explore the potential reasons for this difference, 
attention is drawn to the initial behavior of the accelerated 
particles.
Evolution of energy of an accelerated 
PUI is depicted by the rainbow 
colored lines in Fig.\ref{fig:PUIene}(b) and (c) as a 
function of relative distance from the shock 
and time, respectively. Here, $x_{sh}$ is the shock 
position averaged in the $y$ direction and the 
energy is measured in the simulation (downstream) frame. 
The PUI 
shows typical cycloidal motion up to $t \sim 50$. 
Once it is reflected, the PUI starts gaining energy. 
Beyond $t \sim 60$, there is an increase in energy accompanied 
by periodic fluctuations, indicative of shock drift 
acceleration (SDA) characteristics.
In SDA, an ion gyrating across a shock has relatively small 
gyro radius behind a ramp and large gyro radius in front of 
it so that the ion drifts along the shock surface parallel to 
the motional electric field. The small gyro motion behind a 
ramp is denoted in Fig.\ref{fig:PUIene}(b) as the loop orbit. 
The same type of trajectory is reported by 
\cite{giacalone10} 
(see their Fig.3). However, upon careful observation of the 
trajectory, it is evident that 
there is another acceleration phase 
before $t \sim 60$ and the behavior of the particle during 
this phase differs from that observed afterward.
When first reflected at $x \approx x_{sh}$, 
the particle exhibits a trajectory that is not loop-like, 
indicating that the reflection occurs due to potential. 
The subsequent second reflection is also similar, and such 
multiple reflections due to potential followed by incomplete 
gyro-motion are characteristic of SSA. 

\begin{figure}[ht]
\includegraphics[width=85mm]{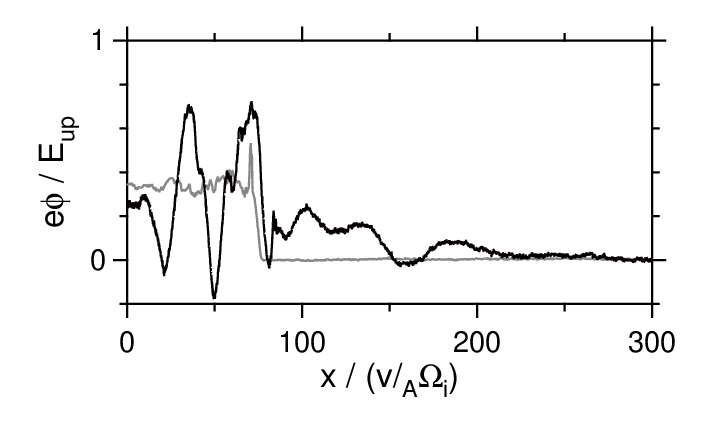}
\caption{\label{fig:potential}
Electrostatic potential for 
$\Theta_{Bn}=50^{\circ}$ (black) and $\Theta_{Bn}=70^{\circ}$ 
(gray) at $t = 50$ along $y \approx 27$. The gray line is 
shifted horizontally in $x$ to align the overshoot position 
with that of the black line.
}
\end{figure}
%
Here, the SSA can work because of the large amplitude 
upstream waves. 
The black line in 
Fig.\ref{fig:potential} represents the profile 
of electrostatic potential normalized to upstream flow 
energy, $e \phi / E_{\rm up}$, along 
$y \approx 27$ at $t = 50$ which is close to the time and position 
in $y$ when the 
PUI in Fig.\ref{fig:PUIene}(b,c) was first reflected at the 
shock. The fluctuation of potential is 
clearly seen both upstream and downstream of the shock ramp 
($x \approx 80$). The upstream fluctuation is compressed 
and amplified at the shock. The potential 
occurs about 0.2 $E_{\rm up}$ at $x \approx 100$ already 
before the ramp. At the ramp, large jump is seen so that 
the potential reaches nearly 0.7 $E_{\rm up}$.
For comparison, the potential at $\Theta_{Bn}=70^{\circ}$ 
at the same time is indicated by the gray line. Here, 
the $x$-value is shifted to align the overshoot position 
almost the same. Without backstreaming ions and upstream 
large amplitude waves, both foot by the reflected PUIs and 
steep ramp can be observed (although not clear in this scale). 
The potential increase at the 
foot is about 0.3 $E_{up}$, and its spatial scale is 
approximately the gyro radius of the reflected PUIs, and 
the gradient of the potential in this part is smaller 
than that of the ramp at $\Theta_{Bn}=50^{\circ}$. The 
potential increase at the ramp is slightly over 0.2 
$E_{\rm up}$, 
and the total potential increase is smaller than that of 
$\Theta_{Bn}=50^{\circ}$. 
Note that the obliquely propagating waves contain 
electrostatic field component. Therefore, the large amplitude 
upstream waves accompany electrostatic potential. When 
$\Theta_{Bn}=50^{\circ}$, the waves are compressed at 
the shock so that the gradient of potential increases. The 
potential jump at the ramp occurs within a few ion inertial 
lengths, although it is larger than the electron inertial 
length which was originally assumed by \cite{lee96}.
Furthermore, the density increase due to the shock compression 
amplifies the potential. These effects can cause the observed 
SSA of the PUI and this additional acceleration occurring 
immediately after the first reflection may enhance the 
production rate of the accelerated PUIs. 
Although we didn't investigate 
all particles, we confirmed that among the particles 
we examined, those with exceptionally high energy 
($E_{\rm PUI}/E_{\rm up} > 60$ at $t=100$) underwent 
the SSA process before the SDA process. 

\section{summary}

We investigated injection process of PUI acceleration at 
a PUI mediated oblique shock by using two-dimensional PIC 
simulation. The evolution of system with sufficiently 
large spatial size in the shock normal direction was 
followed for sufficiently long time ($t> 100$) 
to reproduce accelerated PUIs for the first 
time. Backstreaming PUIs drive large amplitude waves 
through resonant instabilities. The convection of those 
large amplitude waves causes shock surface reformation 
and alters the downstream structure. 
A part of PUIs are accelerated to tens of upstream flow energy with 
the time scale of $t \sim 100$ as observed in 
hybrid simulation by \cite{giacalone21}. 
The initial acceleration occurs through SSA mechanism followed 
by SDA mechanism. The electrostatic potential 
accompanied by the large amplitude upstream waves 
enables the SSA to occur.

\cite{mccomas06} argue that low-energy particles 
are more likely to be accelerated at the terminal shocks in 
the flank of the heliosphere, where shock angles 
are oblique. The results discussed here qualitatively support 
this claim. However, the acceleration of particles in the tens 
of keV range discussed here requires shock angles of 
approximately $60^{\circ}$ or less for injection, whereas 
typical shock angles at the flank may not be so small. 
It may be that the solar wind turbulence advocated by 
\cite{giacalone21,giacalone10} is necessary for injection.

Due to the finite system length in the 
$y-$direction, only limited wave modes are allowed to grow 
in our simulation. In the real system there should be other 
unstable wave modes with different $\theta_{Bk}$. Those waves may 
lead to efficient scattering of the accelerated PUIs, 
which should be confirmed in future.
The accelerated particles in the tens of 
keV ($\sim$tens of upstream flow energy) range observed at 
$\Theta_{Bn}=50^{\circ}$ correspond to the energy range of 
energetic neutral atoms targeted by IMAP (Interstellar Mapping 
and Acceleration Probe) mission, and valuable insights into 
the particle injection site may be obtained from the data 
of its all-sky map.

\section*{Appendix}

The dispersion relation of upstream waves (Fig.\ref{fig:disp}) 
are obtained by numerically solving the linearized Vlasov-Maxwell 
equation system. The detailed derivation of the dispersion 
equation including a beam component at oblique propagation 
is given in Chap.8 of \cite{gary05}. 

We assumed that there are 
three species, background ions, electrons, and beam ions. 
Each distribution function follows a (shifted-)Maxwellian distribution. 
The thermal velocity of electrons is 3.5, corresponding to an 
electron beta of 0.25. The electrons are given a drift velocity 
along the magnetic field to cancel out the current created by 
the beam ions. The background ions include both SWIs and PUIs, 
although we do not distinguish between the two and treat them 
as a single component of background ions at rest with an 
effective thermal velocity of 2.7. This value corresponds to 
an effective ion beta of 15, 
which is the sum of SWI beta and PUI beta. 
We will discuss the validity of this assumption later. 
As shown in Fig.\ref{fig:pui_us_260000}, the drift velocity of the 
beam ions varies depending on the distance from the shock. 
Here, a representative value of 16 is used. The thermal 
velocity of beam ions is set to $v_{tb}=1$. The beam population 
is actually propagating along the magnetic field with finite 
pitch angle so that the distribution function is a ring-beam 
type. This is the reason why the beam population in 
Fig.\ref{fig:pui_us_260000}(i) 
looks anisotropic in the velocity space. 
Since the ring velocity is clearly smaller than the beam velocity, 
we neglect the ring velocity in the dispersion analysis. 
Furthermore, 
a relative beam density of 0.3\% is used. As long as the 
relative density is sufficiently small, the wavelength of 
the excited waves is hardly affected by it. The analysis 
is done in the frame of background ions.

\begin{figure}[ht]
\includegraphics[clip, width=1.0\columnwidth]{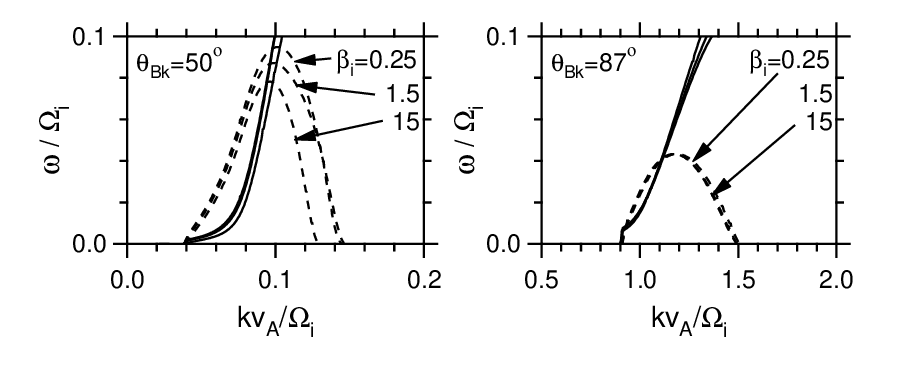}
\caption{\label{fig:appendix}Linear dispersion relation of 
oblique resonant instability driven by BS-PUIs with different 
background ion beta, $\beta_i$. The solid and dashed lines 
indicate real and imaginary parts of frequency, respectively.}
\end{figure}

Fig.\ref{fig:disp} is obtained with the above conditions. 
For $\theta_{Bk}=50^{\circ}$, the maximum growth rate 
occurs at $k \approx 0.096$ and $\omega \approx 0.07$. Since 
the argument of plasma dispersion function, $\zeta_b = (\omega - 
{\bf k \cdot v_b} + \Omega_i) / \sqrt{2} k_{\parallel} v_{tb}$, 
is estimated as $\approx 0.95 (< 1)$, the observed instability 
is a resonant instability. For $\theta_{Bk}=87^{\circ}$, 
the maximum growth rate occurs at $k \approx 1.2$ and 
$\omega \approx 0.065$, leading to $\zeta_b \approx 0.68 
 (< 1)$. Hence, the instability is also a resonant 
instability. Fig.\ref{fig:appendix} shows the dependence 
of the dispersion relation on the beta of the background ions, 
$\beta_i$. 
For both $\theta_{Bk}=50^{\circ}$ and $87^{\circ}$ cases, 
the maximum growth rate and the corresponding wavenumber 
do not very change, while $\beta_i$ varies from 0.25 to 15. 
This indicates that our assumption of considering SWIs and 
incoming PUIs as a single background ion component was valid.

%
%


\section*{acknowledgments}
We thank G. Zank, H. Washimi, and T. Hada for fruitful 
discussion. We also appreciate discussions at the team 
meeting “Energy Partition across Collisionless Shocks” 
supported by the International Space Science Institute 
in Bern, Switzerland. This research was supported by the 
JSPS KAKENHI Grant no 19K03953, 22H01287, 23K22558 (SM) and 
23K03407 (YM). This work was supported by MEXT as “Program 
for Promoting Researches on the Supercomputer Fugaku" 
(Structure and Evolution of the Universe Unraveled by 
Fusion of Simulation and AI; Grant Number JPMXP1020230406) 
and used computational resources of supercomputer Fugaku 
provided by the RIKEN Center for Computational Science 
(Project ID: hp230204)

\bibliography{myaip}{}
\bibliographystyle{aasjournal}

\end{document}